\documentstyle[aps,12pt,amssymb]{revtex} 
 \input psfig
\begin{document}

%
%
\newenvironment{proof}{{\bf Proof:}}{}
\newtheorem{theorem}{Theorem}[section]

\newtheorem{lemma}[theorem]{Lemma}
\newtheorem{proposition}[theorem]{Proposition}
\newtheorem{corollary}[theorem]{Corollary}
\newtheorem{definition}[theorem]{Definition}
\newcommand{\be}{\begin{equation}}
\newcommand{\bea}{\begin{eqnarray}}
\newcommand{\ee}{\end{equation}}
\newcommand{\eea}{\end{eqnarray}}
\newcommand{\ethbar}{\bar{\eth}}
\newcommand{\Lambdabar}{\bar{\Lambda}}
\newcommand{\p}{\partial} 
\newcommand{\zetabar}{\bar{\zeta}}
\newcommand{\omegabar}{\overline{\omega}}
\newcommand{\mbar}{\overline{m}}
\newcommand{\etabar}{\overline{\eta}}
\newcommand{\rhobar}{\overline{\rho}}
\newcommand{\sigmabar}{\overline{\sigma}}
\newcommand{\alphabar}{\overline{\alpha}}
\newcommand{\xibar}{\overline{\xi}}
\newcommand{\th}{\theta}
\newcommand{\Ld}{\Lambda\, }
\newcommand{\ld}{\lambda}
\newcommand{\tLd}{\bar{\Lambda}\, }
\newcommand{\Wbar}{\overline{W}}
\newcommand{\debar}{\bar{\delta}}

\newcommand{\om}{\omega}
\newcommand{\ombar}{\bar{\omega}}
\newcommand{\Om}{\Omega}
\newcommand{\tOm}{\tilde\Omega}
\newcommand{\z}{\zeta}
\newcommand{\zb}{\bar{\zeta}}
\newcommand{\zbar}{\bar{\zeta}}
\newcommand{\Ldp}{\partial_+\Lambda}
\newcommand{\Ldm}{\partial_-\Lambda}
\newcommand{\tLdo}{\partial_1\bar{\Lambda}}
\newcommand{\tLdp}{\partial_+\bar{\Lambda}}
\newtheorem{remark}{Remark}[section]
\newcommand{\Ldo}{\partial_1\Lambda}
\newcommand{\Ldn}{\partial_0\Lambda}
\newcommand{\varid}{\stackrel{\rm def}{=}}
\newcommand{\k}{\vec{k}}
\newcommand{\kcheck}{\check{k}}
\newcommand{\khat}{\hat{k}}
\newcommand{\cI}{({c}_{I}) }
\newcommand{\cII}{({c}_{II}) }
\newcommand{\cIII}{({c}_{III}) }
\newcommand{\g}{{\bf\mbox{g}}}
\newcommand{\f}{{\bf\mbox{f}}}
\newcommand{\scrip}{{\cal I^+}} 
\newcommand{\scrim}{{\cal I^-}}
\newcommand{\scri}{{\cal I}}
\newcommand{\ihat}{\hat{\imath}}
 \newcommand{\jhat}{\hat{\jmath}}
\newcommand{\Hspace}{${\cal H}$-space }
\newcommand{\lone}{\Lambda_1} 
\newcommand{\lonebar}{\Lambdabar_1}
\newcommand{\etatw}{\tilde \eta} 
\newcommand{\zetatw}{\tilde \zeta}
\newcommand{\bareta}{\bar \eta}
\newcommand{\nn}{\nonumber}
\newcommand{\R}{\R}
\def\R{{\bf R}}

\newcommand{\new}{\mbox{\tt\char'134begin\{New\}}}   

\def\endnew{\mbox{\tt\char'134end\{New\}}}  
\newcommand{\cacho}{{\mbox{\bf *CK*}}}
\newcommand{\tere}{{\mbox{\bf *TR*}}}
\newcommand{\mirta}{{\mbox{\bf *MI*}}}
\newcommand{\mnote}[1]{\marginpar{\raggedright\footnotesize\em #1}}
\newcommand{\DELETENOTES}{ 
\renewcommand{\mnote}[1]{}
\renewcommand{\tere}{}
\newcommand{\cacho}{}
\renewcommand{\mirta}{}
\renewcommand{\new}{}
\renewcommand{\endnew}{}
}


\title{Null Surfaces and the Legendre Submanifolds}

\author{  Mirta Iriondo	\and Carlos N. Kozameh \and  Alejandra T. Rojas
	} 
\address{      FaMAF				, 
	       Universidad Nacional de C\'ordoba, 
	       5000 C\'ordoba			, 
	       Argentina			
	} 
      
\maketitle 
\begin{center} Abstract
\end{center}

\begin{abstract}
It is shown that the main variable $Z$ of the Null Surface Formulation
of GR is the generating function of a constrained Lagrange submanifold
that lives on the energy surface $H=0$ and that its level surfaces
$Z=const.$ are Legendre submanifolds on that energy surface. By
globally extending the Lagrange submanifold over $T^*M$ one obtains a
generalized generating function $\widehat Z$ (a generating family). 
Thus, the singularity
structure of the wavefronts can be obtained by studying the projection
map of the Legendre submanifolds to the configuration space. The behaviour of the variable $Z$ at the caustic points is analysed. It is shown  that
except for Minkowski space a single function $Z(x,\z, \zbar)$ cannot
generate the conformal structure of a radiative space-time.
\end{abstract}

\section{Introduction}
The study of null geodesics on a space-time plays an important role in
general relativity. Gravitational lensing, either cosmological, weak or
micro lensing, is the study of the behaviour of null rays on a curved
space-time. Null geodesics are used to define the notion of
singularities on a space-time, to define the null boundary of a
spacetime representing compact objects, etc.

In the last several years a formalism has been developed where null
surfaces play a dynamical role replacing the metric as a basic variable
 \cite{ko-new3,fri-ko-new1,dynamics,i-k-r}. The goal of the Null Surface Formulation of GR, or NSF for
short,  is to introduce a new variable such that from its knowledge one
can obtain {\bf all} the conformal structure of the space-time. Field
equations equivalent to the Einstein's equation then determine the
dynamical evolution of those surfaces.  By casting GR as a theory of
surfaces rather than a theory of fields the NSF provides a completely
new point of view with emphasis on the geometrical character of the
theory. The basic variable is a function $Z(x^a, \z, \zbar)$ with
$x^a$ representing points on the spacetime and $(\z, \zbar)$ parametrizing
the sphere of null directions. At each point on the space-time the
function $Z$ satisfies

\be \label{eq1} 
g^{ab}(x)\partial_aZ(x, \z, \zbar)\partial_bZ(x, \z,\zbar)=0, 
\ee 
and the level surfaces of this function, namely $Z =
const.$ are null hypersurfaces on the space-time. The reader should be
aware that the above construction is done at a local level and that in
general it might not be possible to find a single function satisfying these
conditions on the whole space-time. Weyl curvature induces
self-intersections and caustics on null congruences.  Thus, even if one
locally obtains a smooth hypersurface, extending such surface along the
generators of its null geodesics will fail to be smooth. This
generalized null surface is called a {\em wavefront} and cannot be described
as the level surface of a single function $Z$.

The dynamics of the NSF is imposed as field equations for $Z$ and
another scalar $\Omega$ (the conformal factor) which are equivalent to
the vacuum Einstein equations. The {\bf global regular} solutions (in a
suitable way to be specified in the conclusions) of those equations yield a
radiative space-time, i.e., a space-time representing self interacting
gravitational radiation. We see once again that even from a dynamical
point of view it is very important to study the global behaviour of
this variable, namely the behaviour of wavefronts in General
Relativity.

The purpose of this work is to provide a framework (at a kinematical
level) to discuss global behaviour of the basic variable $Z$. We want
to analyze the singularities of our variable. We want to see under what
circumstances a single function $Z$ suffices to construct the entire
conformal structure, or how many different functions must be given to
cover the space-time.  We only consider a specific class of
space-times, asymptotically flat space-times along future null
directions.  This class of
space-times represents isolated sources that may emit gravitational radiation.

The idea is to study the global solutions of eq.(\ref{eq1}). Note that 
$Z$ can be thought as the action of the Hamilton-Jacobi equation for 
the Hamiltonian $H(x,p) = g^{ab} p_a p_b$. Since equation (\ref{eq1}) can be 
written as 

\[
H(x, \partial_aZ) = 0,
\]
Z is the action of the time-independent Hamilton-Jacobi equation
and the study of the unicity of the solution and its global properties
can be carried out using the tools of analytical mechanics. It is worth
mentioning that the study of the solutions of the Hamilton Jacobi
equations led to the development of the theory of Lagrange
submanifolds on cotangent bundles and the loss of unicity on the
solutions is directly related with the singularities of the projection
map of these submanifolds onto the configuration
space\cite{arnold,marsden}.

In this paper we reintroduce our variable $\widehat{Z}$ as the
generating family of a constrained Lagrange submanifold that lives on
the energy surface $H=0$ and show that its level surfaces are Legendre
submanifolds. Thus, the singularity structure of the wavefronts can be
obtained by studying the projection maps to the configuration space.
We thus, define the caustic set as the points on the Lagrange or
Legendre submanifold with singular projection and the projection of
those points as the caustics.  Since Lagrange and Legendre submanifolds
are smooth surfaces in $T^{*}M$ this work suggests that one can
redefine our variable in a way that is free from the singularities and
self-intersections that are naturally associated with characteristic
wavefronts in GR.

In Section \ref{lag-leg} we introduce the necessary mathematical background 
needed for this work. In this context we also prove that the  hypersurfaces of a constrained Lagrange submanifold defined as the restriction of this  Lagrange submanifold to  the level surfaces of its generating family   are Legendre submanifolds on the energy surface $H=const$.

In Section \ref{toy} we give a specific example to clarify certain
results of the previous section which are technically involved.  We show how
to construct the space-time wavefronts if the generating family is
given and conversely, how to reconstruct the generating family if
several smooth pieces of the wavefront are available. Note that we have
left aside the study of the caustic points on the wavefronts and the
caustic set of the Legendre submanifold that generate those points.
This is done in the next section.

In Section \ref{asyflat} we study the singularity structure of our
variable $Z$ and the main results are found. We show that the caustic
points are obtained by choosing the points were $\eth \ethbar Z$, the
parameter space laplacian of $Z$,  blows up. We also show that at those
points $(Z, \eth Z, \ethbar Z)$ remain finite whereas $\eth^2 Z$ either
vanishes or diverges. Using available singularity theorems we find as a
proposition that except for Minkowski space a single function $Z(x,\z,
\zbar)$ cannot generate the conformal structure of a radiative
space-time. Thus, in order to properly study the global behaviour of
the main variable in the NSF one must abandon the idea of using
a single function on the space-time and instead one has to think of our
variable as a generating family $\widehat Z$ of a Lagrange
submanifold on the cotangent bundle of the space-time. We close this
work with some comments of how to deal with the dynamics of the new
variable.


\section{Lagrange and Legendre Submanifolds.
}
\label{lag-leg}
In this section we review the notions of Lagrange and Legendre
manifolds in a given cotangent bundle $T^{*}M$ of an n-dimensional
manifold $M$. This subject has been fully developed and applied to
different fields in the past twenty years. Exhaustive treatises at a
high mathematical level and/or with beautiful applications to different
areas in physics can be found in the literature
\cite{arnold,marsden,friedrich,golubit,Istewart}.  The review here presented is tailored
to our particular needs and by no means can be considered as a
substitute for the standard references in the field.  

In subsections A and B, we present some definitions and introduce the
concept of {\em constrained Lagrange submanifold} in order to
reinterpret, in the next section, our variable $\widehat Z$ as the
generating family of a Lagrange manifold. Moreover, we prove in
proposition \ref{prop:Legendre} that  the  hypersurfaces of a
constrained Lagrange submanifold defined as the restriction of the
Lagrange submanifold to  the level surfaces of its generating family
are Legendre submanifolds on the  energy surface  $H=const.$

\subsection{Lagrange manifolds}
\label{lagrangian}
Recall that $(P,\omega)$ is a symplectic manifold  if $P$ is an
even-dimensional differentiable manifold and $\omega$ is a closed
nondegenerate differential 2-form on $P$. We consider a particular kind
of submanifolds of $P$  called {\em Lagrange manifolds}.
\begin{definition}
\label{def:lagrange}
Let $(P,\omega)$ be a symplectic manifold of dim $P=2n$,  a manifold $L$ smoothly embedded by a map  $e: L\to P$  is called a {\it Lagrange submanifold}
of $P$ if the pull-back to $L$ of the symplectic form $\omega$  on $P$ by $e$ vanishes on $L$  
$$
  e^{*}\omega=0,
$$
and $L$ is of maximal possible dimension compatible with the symplectic structure $\omega$, i.e. dim $L=n$. 
\end{definition}

For these submanifolds, we introduce functions called {\it generating functions} as follows.
\begin{definition}
\label{def:gener}
Let $(P,\om)$ be a symplectic manifold, $L$ a Lagrange submanifold, and $e: L\to P$ an  embedding. Since locally, $\om=-\mbox{d}\kappa$, then $ e^{*}\om=-d(e^{*}\kappa)=0$, so $e ^{*}\kappa=\mbox{d} S$ for a function $ S:L\to \R$ (locally defined).  We call $ S$ a {\it generating function} for $L$.
\end{definition}

From now on we will restrict ourselves to a particular class of
symplectic manifolds, the cotangent bundle of an n-dimensional
manifold  $M$, denoted by $T^{*}M$. This bundle can be assigned local
coordinates $(q^i,p_i)$ with $(q^i)$ representing points of $M$ and
$p_i$ the local coordinates of the covectors at the point $(q^i)$. In
these local coordinates the closed nondegenerate differential 2-form
$\om$ on $T^{*}M$ can be written as $\om=\mbox{d}q^i\wedge
\mbox{d}p_i$.

If $L$ is a Lagrange submanifold of  $(T^*M,\om)$, then the projection
map $\pi:T^{*}M\to M$ given by $\pi(q^i,p_i)=q^i$ , induces a map
$\overline\pi=\pi\circ e$ called the {\em Lagrange map}. The set of
points where the rank of $\overline{\pi}^*$ drops are called the {\it
singular set} and the image of this set is called the {\it  caustic}.

Notice also that if  $S:M\to {\bf R}$,
then the graph of $dS$, given in local coordinates by
\be
\label{eq:diffeo}
\left \{ (q^i,p_i) \in T^*M: p_i=\frac{\partial S}{\partial q^i}\right\}.
\ee
is a Lagrange submanifold, as can be easily verified. Then, $S$ is the {\it generating function}  of the Lagrange manifold and $\overline{\pi}$ is a diffeomorphism.

The converse is also true, if $\overline{\pi}$ is locally a
diffeomorphism, then $L$ is the graph of $dS$, where $S:M\to {\bf R}$.
In this case $S$ is only locally defined.

Now, consider the hamiltonian system  $(T^{*}M, \om,H)$, where
$H:T^{*}M\to {\bf R} $ is a  hamiltonian function.  The Lagrange
submanifolds  that we are interested on are those that can be
considered as submanifolds of the energy hypersurface $\hat H$ defined
by $H=const$. We shall refer to them as {\it constrained Lagrange
submanifolds}. 

 \begin{definition}
Let $\hat L$ be a Lagrange submanifold of $T^* M$ and $H$ a Hamiltonian
function, we say that $\hat L$ is a constrained Lagrange manifold if
$\hat L\subset \hat H$, where $\hat H$ is an energy surface.
\end{definition}

Constrained Lagrange manifolds have very interesting properties. They
are invariant under the flow of the hamiltonian vector field $X_H$ (\cite{marsden}, Proposition 5.3.32).
This, can be easily proved using the fact that $X_H$ is tangent to the
hypersurface $H= const.$ and that $\hat L$ is of maximal dimension
. Thus, $X_H$ is tangent to $\hat L$ and
the hamiltonian flow preserves $\hat L$.

If $\hat L$ is the graph of $dS$, where $S:M\to {\bf R}$, then its generating function $S$ must satisfy  the time-independent Hamilton-Jacobi equation 
\begin{equation}
\label{jacobi}
H\left(q^j,\frac{\partial S}{\partial q^i}\right)=const,
\ee
i.e., the generating function $S=S(q^i)$ is the action of the Hamilton-Jacobi
equation.  Conversely, a solution of (\ref{jacobi}) locally defines a
constrained Lagrange manifold with a diffeomorphic projection to
$M$.

However, in general (and in the problem we want to address) $\hat L$
will not be globally diffeomorphic to $M$. How do we handle this
situation? Since $\hat L$ is a smooth hypersurface on $T^* M$ contained
in the hamiltonian flow,  at most points the Lagrange projection
will be a diffeomorphism (since the rank of $\overline{\pi}^*$ cannot drop more than
$n-2$  on a set of points of zero measure with respect to the topology of $M$). We thus identify three different regions in  $\hat L$:
\begin{enumerate}
\item[$\bf 1$.]
 smooth open regions of  $\hat L$ diffeomorphic to $M$, and thus
local graphs of a single function $S$,
\item [$\bf 2$.]
points of $\hat L$ where the normal to  $\hat L$ is ``horizontal''
in $T^* M$, i.e., the singular set, which divides {\bf 1} from
\item [$\bf 3$.]
smooth open regions of $\hat L$ that project down to the same
open neighborhood of $M$ (these points of $M$ have more than one
preimage). These regions are generated by several smooth functions $S_i$.   
\end{enumerate}

From the point of view of Hamilton Jacobi theory unique solutions to (\ref{jacobi}) yield regions {\bf 1}, multivalued solutions to (\ref{jacobi}) yield regions {\bf 3} and singularities of the solutions yield the singular set {\bf 2}.

Let us briefly analize how to construct the three different regions from the solutions to the time independent Hamilton-Jacobi equation. 

Since the Hamiltonian flow is tangent to $\hat L$, we need  to solve
Hamilton's canonical equations in order to generate the Lagrange
manifold.  A beautiful method to solve those equations is to find a
generating function of a canonical transformation such that in the new
variables the hamiltonian is independent of the new variables.

We recall that a canonical transformation
is a diffeomorphism in $T^* M$ that  preserves the simplectic structure. Denoting the new variables by $(Q^i, P_j)$. It is then easy to show that 
 \be
\label{eq:exact}
{  p_i}\mbox{d}{  q^i}-{  P_i}\mbox{d}{  Q^i}=\mbox{d}\widehat S({  q^i},{ Q^j}).
\ee
i.e. the difference between the canonical 1-forms associated with the two coordinates is exact. The function $\widehat S$ is called {\it the
generating function} of the canonical transformation. Note that in the
context of Lagrange manifolds this
function $\widehat S$ is not the generating function of  a Lagrange manifold, it is called {\it generating family},  
   and that  in Catastrophe
  Theory  its normal forms are called {\it universal
  unfoldings}\cite{Istewart}. Hence in what follows  we shall refer to    $\widehat
  S$ as the {\it generating family}.

 If the transformation is such that the Hamiltonian  becomes  constant in the new variables, then  the new Hamilton's equations are trivially solved, i.e. ${  P_j}={\beta_j},\; {  Q^i}={\alpha^i}$, where $(\alpha^i, \beta_j)$  are arbitrary constants. Using (\ref{eq:exact})  we get 

\begin{eqnarray*}
-\frac{\partial \widehat S}{\partial Q^i}&=&\beta_i\\
\frac{\partial \widehat S}{\partial q^i}&=&p_i
\end{eqnarray*}
and   this yields the   Hamiltonian flow in the variables $(q^i,p_j)$.

The question is how to find this very specific generating family $\widehat S$. The answer comes from the Hamilton-Jacobi theorem. Given the differential equation

\be
\label{jacobi1}
H\left(q^i, \frac{\partial S}{\partial   q^j}\right)=const
\ee
the complete integral of this equation, $\widehat S=\widehat S(q^i,\alpha_i)$, with $\alpha_i, \;i=1...n, $ arbitrary constants, is a generating function of a canonical transformation.
To see this we set $Q^i=\alpha^i$ and $P_i=\displaystyle{\frac{\partial
\widehat S}{\partial\alpha_i}}$. Then,  if $\displaystyle{\left |\frac{\partial^2
{  \widehat S}}{\partial{  \alpha_i} \partial {  q^j}}\right |}\neq 0$, 
Hamilton's canonical equations can be solved by quadratures. (Jacobi's
Theorem \cite{arnold,landau})

It is then clear that the set $\hat L$ defined by

$$
\hat L=\left \{ (Q^i,P_j)| \; P_j=\beta_j\right \}
$$ 
is a Lagrange manifold for each ${\beta_j}$. In particular, we  can  choose ${  P_j}=0$, or equivalently  
$$
\frac{\partial \widehat S( q^i, \alpha_j)}{\partial  \alpha_i}=0.
$$

Then,  given $\widehat S(q^i,\alpha_l)$ a  complete integral of (\ref{jacobi1}),
with $\alpha_l$ being arbitrary constants for $l=1...n$,   we can
obtain $\hat L$ locally  as the graph of $dS$ by solving first, if it is posible, 
$\alpha_l=\alpha_l(q^i)$ from the equations
 
\be
\label{constrained}
\frac{\partial \widehat S}{\partial \alpha_l}=0
\ee
and then defining $S(q^i)=\widehat S(q^i, \alpha_j(q^i))$.  This can be guaranteed if the rank of the system (\ref{constrained}) is $r=n$ in the variables $\alpha_l$. The Lagrange submanifold $\hat L$ is described by setting
$$
 p_i=\frac{\partial \widehat S}{\partial q^i},
$$
and the {\em Lagrange map} $\bar \pi$ is a diffeomorphism (i.e. the rank of $\overline\pi^*=n$).

If $r=k<n$ then   there exists  $\alpha_J=\alpha_J(q^i)$, for  $J=1...k$ and $\widehat S=\widehat S(q^i,\alpha_I)$, for $I=k+1...n$. In this case the rank of $\overline\pi^*$ drops at some points and this  can be related to the presence of these parameters $\alpha_I$, for $I=k+1...n$.


Then, the solution  of (\ref{jacobi1}), $\widehat S(q^i, \alpha_I)$, defines a Lagrange submanifold $\hat L\subset\hat H$    embedded into $T^{*}M$ by setting:
\begin{equation}
\label{eq:pi}
p_i=\frac{\partial \widehat S}{\partial q^i} \hspace{1cm} 1\leq i\leq n,
\end{equation}
and  imposing the constraints 
 \be
\label{eq:const1}
0=\frac{\partial \widehat S}{\partial \alpha_I}.
\ee
Since the rank of (\ref{eq:const1}) is $n-k$ in the $q^I$ variables, then there exist  $q^I=q^I(\alpha_I, q^J)$, and $\hat L$ and $\overline{\pi}(\hat L)$ are  parametrized by $(\alpha_I, q^J)$. The derivative  $\overline\pi^*$ can be written  as 
$$
\left (\begin{array}{ll}
\displaystyle{\frac{\partial q^I}{\partial \alpha_I}}
&\displaystyle{\frac{\partial q^I}{\partial q_J }}\\
0&I
\end{array}\right ),
$$
where $I$ is the identity matrix $k\times k$, therefore it  is clear that the rank of  $\overline{\pi}^*\geq k$  and it shall be strictly less than $n $ when 
$$
\left |\frac{\partial q^I}{\partial \alpha_I}\right |=0.
$$
The set of singular points and thus the caustic set shall be isolated points, curves or in general a set of points of zero measure with respect to the topology of $M$, since  the  rank of $\overline\pi^*$ cannot drop more than $n-2$. This assertion can be easily understood since there are two vector in $T^*\hat L$ that can not vanish under the projection, one is $X_H$ and the other is the dual of $d\widehat S$.
Notice  that we can nevertheless write $\alpha_I=\alpha_I(q^i)$ if we allow $\alpha_I$ to be  multivalued functions. As a consequence we obtain multivalued generating functions $S_i$.

\subsection{Legendre manifolds}
Odd-dimensional manifolds do not admit a {\em symplectic structure}.
The analogue of a symplectic structure for odd-dimensional manifolds is a {\em  contact structure}.

\begin{definition}
A {\it contact manifold} is a pair $(\hat P, \hat \om)$, consisting of an odd-dimensional manifold $\hat P$ and a closed 2-form $\hat \om$ of maximal rank on this manifold. An exact contact manifold $(\hat P,\hat\kappa)$ consists of a $(2n-1)$-dimensional manifold $\hat P$ and a 1-form $\hat \kappa$ on $\hat P$ such that $\hat\om=-d\hat\kappa$ is of maximal rank  on $\hat P$.
\end{definition}

Moreover, we can define a submanifold analogous to a Lagrange manifold, $N$ of $\hat P$ called a {\em Legendre submanifold}.  
\begin{definition}
Let $(\hat P,\hat \kappa)$ be a contact manifold  of dimension $2n-1$, a $(n-1)$-dimensional manifold  $N$ such that
$$
\hat e^{*}\hat\kappa=0,
$$ 
with $\hat e:N\to \hat P$ an embedding, is called a {\it Legendre submanifold} of $\hat P$.
\label{def:Legendre}
\end{definition}

Now, consider the hamiltonian system  $(T^{*}M, \om,H)$,  the next proposition ensures that we can find, in a natural way, a {\it  contact submanifold of $T^{*}M$}.
\begin{proposition}
\label{prop:energy}
Let $(T^{*}M,\om,H)$ be a Hamiltonian system and $\hat H$ a regular energy surface, defined by $H=const$. Then $(\hat H,i^{*}\om)$ is a contact manifold, where $ i:\hat H\to T^{*}M$ is an inclusion (\cite{marsden}, Proposition 5.1.7).  
\end{proposition} 
Thus the Legendre submanifolds we will consider are those that are submanifolds of   the contact manifold $(\hat H,i^{*}\om)$. Moreover they are hypersurfaces of {\it constrained
Lagrange manifold} of a given Hamiltonian system and  the projection map $\pi:T^*M\to M$ will induce a map $\hat\pi$ defined as $\hat \pi=\pi\circ \hat e$ and called {\em Legendre map}.

 The next proposition
gives a description of this kind of manifolds.

\begin{proposition}
\label{prop:Legendre}
Let $\hat L$ be a constrained Lagrange submanifold of the Hamiltonian
system $(T^*M,\om,H)$ and $\widehat S$ its  generating family, i.e. a
solution of the time independent Hamilton-Jacobi equation . Then the
hypersurface $\hat N$ of $\hat L$, defined as the restriction of  $\hat L$ to   $\widehat S=const$, is a Legendre submanifold of $\hat
H$.
\end{proposition}
\begin{proof}
Given a Hamiltonian system, the  Proposition \ref{prop:energy} ensures that $(\hat H,i^*\kappa)$ is a contact manifold and since $\hat L$ is a constrained Lagrange manifold, the  generating function $\widehat S(q^i,\alpha_I)$ of  $\hat L$ satisfies $H(q^i,\partial_jS)=const$. Then $\widehat S$  defines a Legendre submanifold $\hat N$ of $\hat H$ by setting
\be
\label{eq:pi1}
p_i=\frac{\partial \widehat S}{\partial q^i} \hspace{1cm} 1\leq i\leq n,
\ee 
imposing the constraints
\be
\label{eq:const2}
\widehat S=const. \quad \mbox{and}\quad
\frac{\partial \widehat S}{\partial \alpha_I} = 0,
\ee
 and requiring that the rank of (\ref{eq:const2}) shall be  $n-k+1$ in the $q^I$ variables.

Recall   that  (\ref{eq:const2}) is an algebraic non-linear system of equations, then if we define the function   $G=(\widehat S,\frac{\partial \widehat S}{\partial \alpha_I})$, a solution of (\ref{eq:const2}) satisfies $G={\bf 0}$, that is, it belongs to the kernel of the map $G$.  Therefore demanding that the rank of the derivative of  $G$ to be $n-k+1$ in the variables $q^I$ and in one of the $q^J$ variables, the implicit function Theorem  guarantees that $q^i=q^i(q^j,\alpha_I)$ for $i\in I+1$ and $j\in J-1$. 

Observe that the Legendre manifold constructed in this way becomes an hypersurface of $\hat L$ and that both are submanifold of the energy surface $\hat H$.$\Box$
\end{proof}

The image of the Legendre map is called the {\it wavefront} and the image of the constrained Lagrange manifold can be considered as a wavefront family.  As in the case of the Lagrange manifolds, 
the set of points where the rank of $\hat{\pi}^*$ drops are called the {\it singular set} and the image of this set is called the {\it  caustic}. If the singular set of $\hat L$ is known then intersection of this set with $\widehat S$ = const. yields the singular set of the associated Legendre submanifold.

\section{Examples: A toy model }
\label{toy}
To clarify the concepts of constrained Lagrange and Legendre
submanifolds as its level surfaces we present in this section a
specific example. Using a particular generating family we describe the
singularity structure of the wavefront. We also show  how many pieces
of smooth wavefronts are available on the configuration space for a
given a generating family. (Conversely, the Legendre manifold can be
reconstructed if several pieces of smooth wavefronts are given.) We
leave aside the issue of caustic points and caustics on the wavefronts
which are analyzed in the next section.

Since we are interested in null surfaces on a Lorentzian manifold
$(M,g_{ij})$ as Legendre submanifolds of an energy surface in
$T^{*}M$,  we take $H=\frac{1}{2} g^{ij}(q^k)p_ip_j$ as our Hamiltonian
function and consider the hypersurface $H=0$.

For simplicity, assume that $n=3$  and $g_{ij}= \mbox{diag}(1,-1,-1)$. Then the Hamilton-Jacobi equation associated with this $H$ is
\be
\label{eq:eikonal}
\left(\frac{\partial S}{\partial q_1}\right)^2-\left(\frac{\partial S}{\partial q_2}\right)^2-\left(\frac{\partial S}{\partial q_3}\right)^2=0,
\ee
i.e. the eikonal equation. The complete integral of this equation is

$$
\widehat S=\alpha_0+\sum_{i=1}^3 \alpha_i q_i,
$$
where $\alpha_i$ satisfy
$$
\alpha_1{}^2-\alpha_2{}^2-\alpha_3{}^2=0.
$$
and $\alpha_0=\widehat S_{|q^i=0}$. Hence the complete solution   can be written as
$$
\widehat S(q^i,\alpha_I)=\widehat S(0,\alpha_I)+\alpha_1q_1+\alpha_1\sqrt{1-
\left(\frac{\alpha_2}{\alpha_1}\right)^2}q_2  +\alpha_2 q_3,
$$
with $I=1,2$. The appearence  of caustics shall depend on the choise of the parameters $\alpha_I$ and $\widehat S(0,\alpha_I)$ or equivalently on the initial value of the integral curves of $X_H$. Choosing $\widehat S(0,\alpha_I)= F(\alpha_I)$, were $F$ are the germs of the normal forms of the  generating  functions of Lagrange manifold   \cite{arnold,Istewart}, we obtain the generating family for the singularities type $A_2, A_3$ o $A_4$.  For example a cusp can be obtained if we choose $ \widehat S(0,\alpha_I)=-\displaystyle{\frac{\alpha_1{}^4}{2}}$ and $\alpha_2= \displaystyle{\frac{\alpha_1{}^2}{2}}$. Note that the universal unfolding of a cusp is ( see \cite{Istewart})
$$
\widehat S(\alpha,x,y)=\pm \alpha{}^4 +x\alpha^2+y\alpha
$$
Then, the functions 
\begin{eqnarray*}
\widehat S(q_1,q_2,q_3,\alpha)&=&-\frac{\alpha^4}{4}+q_1\alpha +\frac{1}{2}q_2\alpha (4-\alpha^2)^{\frac{1}{2}}+\frac{1}{2}q_3\alpha^2\\
\frac{\partial\widehat S}{\partial \alpha}(q_1,q_2,q_3,\alpha)&=&-\alpha^3+q_1+q_3\alpha+\frac{1}{2}q_2(4-\alpha^2)^{\frac{1}{2}}
-\frac{1}{2} q_2\alpha^2(4-\alpha^2)^{-\frac{1}{2}}
\end{eqnarray*}
 define a {\it constrained Lagrange manifold}. Note that the Taylor expansion of  $\widehat S(q_1,q_2,q_3,\alpha)$ around $\alpha = 0$ is precisely the universal unfolding given above. The explicit construction follows.

From the equation
\be
\label{eq:const3}
\frac{\partial \widehat S}{\partial \alpha}=-\alpha^3+q_1+q_3\alpha+\frac{1}{2}q_2(4-\alpha^2)^{\frac{1}{2}}
-\frac{1}{2} q_2\alpha^2(4-\alpha^2)^{-\frac{1}{2}}=0,
\ee
we  trivially obtain
$$
q_1=\alpha^3-q_3\alpha-\frac{1}{2}q_2(4-\alpha^2)^{\frac{1}{2}}+\frac{1}{2} q_2\alpha^2(4-\alpha^2)^{-\frac{1}{2}},
$$ 
and by (\ref{eq:pi}) we write  $p_1=\alpha$,  $p_2=\displaystyle\frac{1}{2}\alpha(4-\alpha^2)^{\frac{1}{2}}$ and $p_3=\displaystyle\frac{\alpha^2}{2}$.   

 The map $e:{\bf R}^3\to T^{*}M$
\begin{eqnarray*}
e(q_2,q_3,\alpha)&=&\bigg (\alpha^3-q_3\alpha-\frac{1}{2}q_2(4-\alpha^2)^{\frac{1}{2}}+\frac{1}{2} q_2\alpha^2(4-\alpha^2)^{-\frac{1}{2}},q_2,q_3,\\
& &\qquad \alpha,\displaystyle\frac{1}{2}\alpha(4-\alpha^2)^{\frac{1}{2}}, \displaystyle\frac{\alpha^2}{2}\bigg )
\end{eqnarray*}
is an embedding and  since $\widehat S$ satisfies (\ref{eq:eikonal}) this surface is  in $H=0$. The Lagrange map becomes
\be
\label{cusp}
\bar \pi^i(q_2,q_3,\alpha)=(\alpha^3-q_3\alpha-\frac{1}{2}q_2(4-\alpha^2)^{\frac{1}{2}}+\frac{1}{2} q_2\alpha^2(4-\alpha^2)^{-\frac{1}{2}},q_2,q_3),
\ee
clearly in a neighbourhood of $\alpha=0$

$$
\bar \pi^i(q_2,q_3,\alpha)=(\alpha^3-q_3\alpha-q_2(1-\frac{3}{8}\alpha^2),q_2,q_3),
$$
and this yields a cusp.
 
The map (\ref{cusp}) is not a diffeomorphism, since the equation 
$$
J=\mbox{det}(\bar \pi^*)=-3\alpha^2+q_3+q_2\alpha(4-\alpha^2)^{-\frac{3}{2}}(\alpha^2-6)=0,
$$
gives us the points where the map losses its rank. Solving this equation we obtain 

\begin{eqnarray*}
q_3&=&f(q_2,\alpha)\\
&=&3\alpha^2-q_2\alpha(4-\alpha^2)^{-\frac{3}{2}}(\alpha^2-6).
\end{eqnarray*}
Then the caustic set  is the image of $e(q_2,f(q_2,\alpha),\alpha)$ and the  caustic is the image  of $ \bar\pi^i(q_2,f(q_2,\alpha),\alpha)$. The last  is shown in FIG \ref{fig1}. 

\begin{remark}
\label{multiple-Lag}
Observe that the projection of the surface $J=0$ into configuration space (i.e. the caustic) divides regions on  which the Lagrange map is a diffeomorphism, for example  the region given by $\alpha >0, q_2>0$ and $q_3<0$,  from other regions  where the Lagrange map in not injective (more than one preimage).   In the first region we may locally write   $\alpha=\alpha(q^i)$ from equation (\ref{eq:const3}) and the generating function  $S: M\to {\bf R}$. 
On the other hand, in the regions of non-injectivity, we obtain  more than one function $\alpha_i$  as solution of the equation (\ref{eq:const3}) which in turn implies that we get several functions $S_i:M\to {\bf R}.$
\end{remark}
\begin{figure}
\hspace{-1.2in}\psfig{file=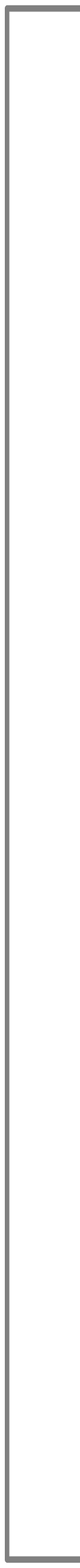,height=3in,width=2.5in}
\caption{ The caustic}
\label{fig1}
\end{figure}

The corresponding Legendre submanifold is then constructed as a level surface of the Lagrange manifold presented above.

As before we write $p_1=\alpha$, $p_2=\displaystyle\frac{1}{2}\alpha(4-\alpha^2)^{\frac{1}{2}}$ and $p_3=\displaystyle\frac{\alpha^2}{2}$. 
  From the   system
\begin{eqnarray*}
S=-\frac{\alpha^4}{4}+q_1\alpha+\frac{1}{2}q_3\alpha^2+\frac{1}{2}q_2\alpha (4-\alpha^2)^{\frac{1}{2}}&=&0\\
 \frac{\partial S}{\partial \alpha}=-\alpha^3+q_1+q_3\alpha+\frac{1}{2}q_2(4-\alpha^2)^{\frac{1}{2}}
-\frac{1}{2} q_2\alpha^2(4-\alpha^2)^{-\frac{1}{2}}&=&0,
\end{eqnarray*}
we obtain   
$$
q_1=\frac{1}{2}\alpha^3-2q_2(4-\alpha^2)^{-\frac{1}{2}},\hspace{.5cm}
q_3=\frac{3}{2}\alpha^2+q_2\alpha(4-\alpha^2)^{-\frac{1}{2}}
$$

The map $\hat e:{\bf R}^2\to T^*M$
\begin{eqnarray*}
\hat e (q_2,\alpha) &=&\bigg (\frac{1}{2}\alpha^3-2q_2(4-\alpha^2)^{-\frac{1}{2}},q_2,\frac{3}{2}\alpha^2+q_2\alpha(4-\alpha^2)^{-\frac{1}{2}}
,\\
&&\qquad \alpha,\frac{1}{2}\alpha(4-\alpha^2)^{\frac{1}{2}}, \frac{\alpha^2}{2}\bigg )
\end{eqnarray*}
is an embedding and it defines a Legendre submanifold. The Legendre map 
$$
\hat \pi^i(q_2,\alpha)=(\frac{1}{2}\alpha^3-2q_2(4-\alpha^2)^{-\frac{1}{2}},q_2,\frac{3}{2}\alpha^2+q_2\alpha(4-\alpha^2)^{-\frac{1}{2}})
$$
describes a wavefront, i.e. $\hat\pi(\hat N)$. It is  shown in  FIG \ref{fig2}a.

\begin{figure}
\hspace{-2.6in}\psfig{file=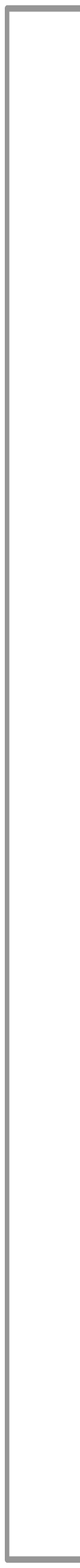,height=3in,width=2.5in}
\hspace{.6in}\psfig{file=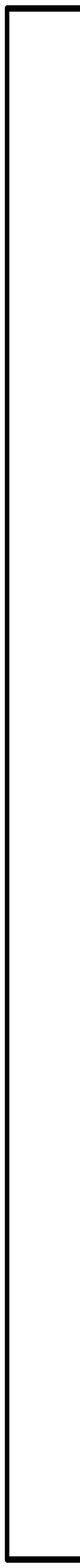,height=3in,width=2.5in}
\caption{(a) The wavefront and (b) $J=0$}
\label{fig2}
\end{figure}
 The derivative of the Legendre map is calculated as
$$
\hat\pi^*=\left (\begin{array}{cc}
 -\alpha f(\alpha,q_2)& -2(4-\alpha^2)^{-\frac{1}{2}}\\
0&1\\
2 f(\alpha,q_2)&\alpha(4-\alpha^2)^{-\frac{1}{2}}
\end{array}\right ),
$$
where $f(\alpha,q_2)=-\displaystyle{\frac{4q_2+3\alpha(4-\alpha^2)^{\frac{3}{2}}}{2(4-s^2)^{\frac{3}{2}}}}$. Clearly, the map $\hat \pi$ is not a diffeomorphism since the rank of its differential drops when 
$$
4q_2+3\alpha(4-\alpha^2)^{\frac{3}{2}} =0,
$$
i.e. on the curve $(\alpha,q_2(\alpha))$, $\alpha\in [-2,2]$. This region (the caustic set) is depicted in  FIG \ref{fig2}b.
The caustic defined by these points (i.e. the image of the curve given above under $\hat \pi$) is given by 
$$
q_1=-\alpha(\alpha^2-3),\qquad   q_2=-\frac{3\alpha}{4}(4-\alpha^2)^{\frac{3}{2}}\qquad\mbox{and}\qquad  q_3=\frac{3\alpha^2}{4}(\alpha^2-2),\quad \alpha\in [-2,2].
$$
 and it is drawn in FIG \ref{fig4}a.
\begin{figure}
\hspace{-2.6in}\psfig{file=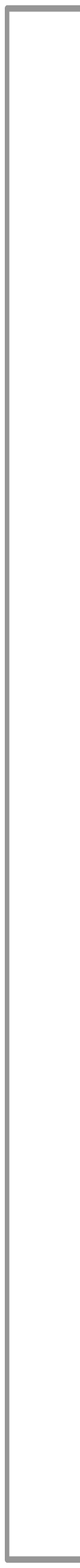,height=3in,width=2.5in}
\hspace{.6in}\psfig{file=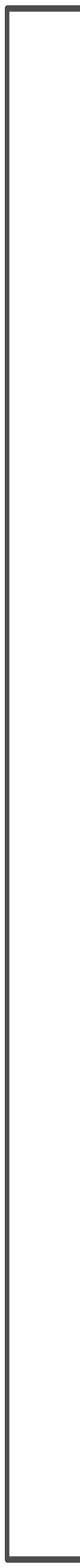,height=3in,width=2.5in}
\caption{(a) The caustic and (b) the null geodesics}
\label{fig4}
\end{figure}

 This caustic, together with some integral curves of the non-vanishing null vector $l^i=\displaystyle{\frac{\partial \hat\pi^i}{\partial q_2}=\left(-2(4-\alpha^2)^{-\frac{1}{2}},1,\alpha (4-\alpha^2)^{-\frac {1}{2}}\right)}$
are shown in FIG \ref{fig4}b. 
Note that the caustic is the envelope of the null geodesics with tangent vector $l^i$ and that  the tangent vectors to the curves given $q_2=const$ vanish on the caustic. The vanishing of these tangent vectors $M^i=\displaystyle{\frac{\partial\hat \pi^i}{\partial \alpha}}$  can be seen in FIG \ref{fig6}, where the caustic and some $q_2=const$ curves (surfaces in higher dimensions) are shown.

\begin{figure}
\hspace{-1.2in}\psfig{file=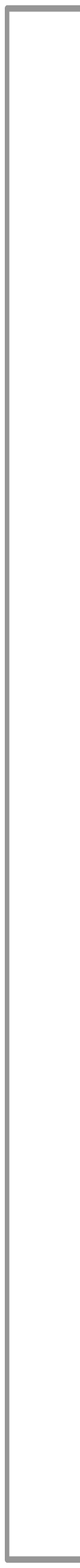,height=3in,width=2.5in}
\caption{ The curves $q_2=const$}
\label{fig6}
\end{figure}

Observe that in the region given by $-2\leq \alpha\leq 2$ and $ q_2 > 4$ the Legendre map is a diffeomorphism (see  FIG \ref{fig2}). Thus, for a sufficiently large $q_2$, the function $\widehat S|_{q_2=const}$ defines a submanifold $\hat Q$ diffeomorphic to $Q\approx {\bf R}\times[-A,A]$, being $A$ a positive and large constant (see FIG \ref{fig7}). 
\begin{figure}
\hspace{-1.2in}\psfig{file=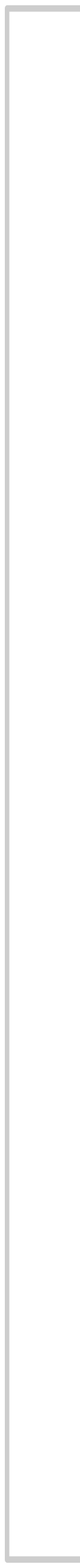,height=3in,width=2.5in}
\caption{ A diffeomorphic region}
\label{fig7}
\end{figure}

From this example we conclude that:
 
\begin{remark}
\label{multiple}
 Since there are regions outside the caustic set where  the generating function   $S(q_i)$  is multiple-valued, we have  several pieces of a smooth wavefront. Conversely if we have several smooth pieces of a wavefront, using  (\ref{eq:pi1}), we can  reconstruct the Legendre manifold  except at the caustic  points. 
\end{remark}

\section{The function $Z$ in asymptotically flat space-time}
\label{asyflat}

For simplicity we will consider asymptotically flat space times with a null boundary that represents the end points of future directed null geodesics\cite{hawking}. Those space times represent compact objects that can emit gravitational radiation.

To define our variable $Z$ we consider the intersection of the future
null cone from $x^a$ with the null boundary $\scrip$. Introducing Bondi coordinates $(u, \z, \zb)$ on ${\cal
I}^+$ (with $u$ representing a Killing time and $(\z, \zb)$ being
stereographic coordinates on the unit sphere) this intersection is
locally given by the equation

\be
u=Z(x^a,\z,\zbar).
\label{eq:Z1}
\ee
Thus, for fixed values of $x^a$, the function $Z$ yields the parametric description of the light cone cuts of $\scrip$.

On Minkowski spacetime the l.c. cuts adopt a very simple form, $u =
x^al_a$ with $l_a$ a null covector constructed from the spherical
harmonics $Y_{0,0}$, $Y_{1,-1}$, $Y_{1,0}$, $Y_{1,1}$.

In general a cut is a complicated surface with caustics,
self-intersections, etc. However, it can be shown that for regular
space-times the index number is always one. Therefore, a
l.c. cut (as complicated as it might be) is always a continuous
deformation of the sphere of null directions above each point $x^a$. It
can also be shown that generically the l.c. cuts can only have two
kinds of singularities, cusps and swallowtails since they represent the
projection of 2-dim Legendre submanifolds on $\scrip$\cite{klr-91}.

A second meaning can be assigned to our variable  $Z(x^a, \z, \zbar)$. Fixing
a point $(u,\z,\zbar)$ of $\scrip$, the collection of interior points
$x^a$ that satisfy

\be
\label{eq:Z2}
Z(x^a,\z,\zbar)=u=const,
\ee
form the past null cone of $(u,\z,\zbar)$. Moreover, from knowledge of
$Z$ we can construct a null coordinate system as follows.

Starting with our variable  and taking $(\z,\zb)$ derivatives of $Z(x^a,\zeta,\bar\zeta)$ we construct the following set of scalars,

\be
        \theta^i(x^a,\zeta,\bar\zeta)
   \equiv
        (\theta^0, \theta^+, \theta^-, \theta^1)
   \equiv
        (u,\omega,\bar\omega,R)
   \equiv
        (Z,\eth Z,\ethbar Z,\eth \ethbar Z).          \label{2.2}
  \ee
For fixed values of $(\zeta,\bar\zeta)$ they define a coordinate system with
the following geometric meaning, 

\begin{itemize}
\item u = const. denotes the past null cone from $(u,\z,\zbar)$. 
\item $(\omega,\bar\omega)$ = const. single out a null geodesic on that 
      surface.
\item $R$ = const. identifies a point on that geodesic. 
\end{itemize}

However, one knows that null cones can develop caustics and
singularities. One also knows that past those singularities the null
cone is no longer smooth (it is called a wavefront) and thus, a null
coordinate system like the one above breaks down past those singular
points. Since the main goal of the NSF is to replace the metric with a
function $Z$ such that its level surfaces are past null cones from
$\scrip$, we immediately face a non trivial problem: if the null cones
develop self-intersections and singularities that cannot be analyzed
with a single function $Z$ then the construction given above is only
valid on a neighborhood of $\scrip$. However, we also know that null
wave fronts are projections of Legendre submanifolds that live on
$T^*(M)$. It would then appear that the best way to deal with this lack
of smoothness is to think of our variable as the generating family
$\widehat Z$ of a constrained lagrange submanifold. An outline of this
construction is presented below.

As was done before we assume that $Z$ is a solution of the
equation

\begin{equation}\label{eq:H}
H(x^a,\partial_b Z) = g^{ab}Z,_aZ,_b=0.
\end{equation}
with $g^{ab}$ a metric that is asymptotically flat.
 
In a neighborhood of $\scrip$ the solution to this equation yields a single function $Z$ and its level surfaces $Z=const.$ describe the past
null cones from points at the null boundary. This follows from the fact
that the (unphysical) metric near $\scrip$ is ``almost'' conformally
flat and thus the past null cones are free from caustics and
singularities.

Since $Z(x^a,\z_0,\zbar_0)$ is a smooth function on this region, we can choose $S=Z(x^a,\z_0,\zbar_0)$ as the generating function of a constrained Lagrange manifold  $\hat L$
\begin{equation}\label{eq:Lagrange}
N=\left \{ (x^a,p_b=\frac{\partial Z}{\partial x^b}) : e^{*} \kappa=\mbox{d}Z
\right \}.
\end{equation}
Note that the manifold described above is equivalent to (\ref{eq:diffeo}) and the Proposition \ref{prop:Legendre}  ensures that the surface $\hat N$ defined by  $ Z=const$ is a  Legendre submanifold of the energy surface given by $ H=0$, i.e. 
$$ 
\hat N=\left \{ (x^a,p_b=\frac{\partial Z}{\partial x^b}) : \hat e^{*} \hat\kappa=\mbox{d}(e^{*}Z)=0
\right \}.
$$

Thus, we have constructed a  constrained Legendre submanifold $\hat N$ and  a constrained Lagrange submanifold $\hat L$ of $\hat H$ using our fundamental variable $Z(x^a,\z_0,\zbar_0)$.

The idea is to extend this construction to regions were caustics
develop.  As we mentioned before, in these regions the Lagrange
submanifold is not diffeomorphic to its  projection. We will thus
assume that a solution of  $H(x^a,\partial_b Z)=0$  can be written as
$\widehat Z=\widehat Z(x^a,\z_0,\zbar_0,\alpha_I)$ with $I=1,2$. The
function $\widehat Z$ depends, at most, on two parameters since the
rank of the  projection map cannot drop more than two\cite{arnold}.
The constrained Lagrange submanifold $\hat L$ as described in section \ref{lagrangian} is
given by

$$
 p_a=\frac{\partial \widehat Z}{\partial x^a}.
$$
together with the constraint
\be
\label{const5}
\frac{\partial \widehat Z}{\partial \alpha_I}=0.
\ee
 Observe that if we can solve (\ref{const5}) uniquely for $\alpha_I$, i.e. $\alpha_I=\alpha_I(x^b)$, then $\widehat Z=Z$ and we are back in the previous diffeomorphic region. In general, one will obtain multivalued solutions of (\ref{const5}). Inserting the different solutions of $\alpha_I$ into $\widehat Z(x^a,\z_0,\zbar_0,\alpha_I)$ one obtains a multiple-valued function $Z(x^a,\z_0,\zbar_0)$. The Legendre submanifold is obtained by setting $\widehat Z=const$. Conversely, if several functions $Z_i$ are given, one can reconstruct the Lagrange submanifold by imposing (\ref{eq:Lagrange}) on the different $Z$'s. The construction defines the lagrange submanifold except for the caustic set.
 
Finally, we would like to determine under what circumstances it is possible to find a single function $Z$ that would yield for us a global coordinate system
$(u,R,w,\bar w)$ on an asymptotically flat space time. In other words we want to know if there exists space times that are diffeomorphic to the corresponding Lagrange manifolds.  At the same time we would like to know when and how this coordinate system breaks down due to the presence of conjugate points. We are therefore interested in describing the relationship between  our fundamental variable $Z$ and the loss of the rank of the derivative of the Legendre map $\hat \pi$, i.e. we want to described the singular set in terms of $Z$. 

When the Lagrange manifold is a  constrained one, the loss of rank of the Lagrange map indicates the non-existence of global solutions of the Hamilton-Jacobi equation and  the loss of rank of the associated  Legendre 
map is related to the {\it existence of conjugate points of a congruence of null geodesics}. 

In order to clarify this assertion, we  consider the local description of the wavefront (the projection of the Legendre manifold). We assume that the wavefront is locally described by
$$
x^a=f^a(u_0,s, w,\bar w, \z_0, \zbar_0).
$$
with $s$ an affine length. The vectors
$$
L^a=\frac{\partial f^a}{\partial s},\qquad M^a=\frac{\partial f^a}{\partial w}\qquad\mbox{and}\qquad\bar M^a=\frac{\partial f^a}{\partial \bar w}.
$$
are tangent to the wavefront.  $L^a$ is directed along the null geodesics whereas $M^a$ and $\bar M^a$ are geodesic deviation vectors.

The derivative of the Legendre map losses its rank when these three vectors become linearly dependent. This dependence is related to the existence of conjugate points on the congruence of null geodesic with apex at $\scrip$ and null tangent vector $L^a$ as follows.

We introduce the parallelly propagated null triad $\{l^a,m^a,\bar m^a\}$, satisfying

\be
\label{eq:tetrad}
\left .\begin{array}{ll}
l^am_a&=0\\
m^a\bar m_a&=-1\\
l^a\nabla_am^b&=0.
\end{array}\right \}
\ee

In terms of this  triad 
\be \label{triad}
L^a=l^a,\quad M^a=\xi m^a+\bar \eta\bar m^a, \quad \bar M^a=\bar\xi \bar m^a+\eta m^a,
\ee
therefore this set of vectors  becomes linearly dependent when
\be
\label{eq:singular}
\left |\begin{array}{cc}
            \xi&\eta\\
             \bar \eta& \bar\xi
         \end{array}
\right |=(\xi\bar\xi-\eta\bar\eta)=0.
\ee
On the other hand, this quantity  is related to the divergence $\rho$, and the shear $\sigma$ of the congruence with apex in $\scrip$. To see this , consider the optical parameters
$$
\rho=m^a\bar m^b\nabla_a l_b\qquad \mbox{and}\qquad\sigma=m^a m^b\nabla_a l_b.
$$
Using eq.(\ref{triad}) together with the fact that $M^a$ is Lie propagated along the null direction $L^a$ we get(\cite{dynamics})  
$$
\sigma=\frac{\bar\eta^2}{A}D\left(\frac{\bar\xi}{\bar\eta}\right)\qquad 
\rho=\frac{DA}{2A}
$$
with $A=(\xi\bar\xi-\eta\bar\eta)$ and where we have used the fact that $\rho$ is real.

Hence, at the points where the Legendre map looses its rank, the divergence of the congruence becomes unbounded, i.e. $\lim \rho_{s\to s_0}=\infty$, where $s$ is the affine parameter and $s_0$ corresponds to a conjugate point.

{\it Summarizing, the loss of rank of the Legendre map, i.e. the development of a caustic is directly related to the existence of a conjugate point on the null congruence with apex at $\scrip$.} 

Since $Z$ is a unique function near $\scrip$, $(u,R,w,\bar w)$ is a well behaved coordinate system in that region. The question is what happens to our coordinates as we approach a generic conjugate point. 

\begin{lemma}
Let $Z(x^a, \z, \zbar)$ be our basic variable near $\scrip$ such that $Z=0$ describes the past null cone from $(u, \z, \zbar)$. Then at the first conjugate point of this null congruence the scalar $R=\eth\ethbar Z$ goes to $-\infty$.
\label{lemma:R}
\end{lemma}
\begin{proof}
Given a null geodesic labeled by $(u, w,\bar w, \z, \zbar)$ we introduce an affine length $s$ and two null congruences that contain this geodesic; 1) the future null cone with apex at $s_0$ and 2) the past null cone with apex at $(u, \z, \zbar)$. 
 The divergence of the first congruence at $\scrip$ is related to the value of the scalar $R(s_0)$ in the following way\cite{ko-cut-new}. The divergence of the generating vectors of the first cone is defined by

$$
\rho_1=m^a\bar m^b\nabla_a F_b.
$$
where $F=F(\Omega,u,x^i)$ and  $F=0$ describes the null cone. Near $\scri$  the function $F$ can be written as 
$$
F=F^0(u,x^i)+\Omega F^1(u,x^i)+{\cal O}(\Omega^2).
$$
where $F^0=u-Z$. Then 

\begin{eqnarray} 
\label{div}
\rho_1(s_0,\scrip) &=&m^a\bar m^b\nabla_a(u-Z)\nonumber\\
&=&\rho_B - R(s_0).
\end{eqnarray}
where $\rho_B$ is the divergence of a Bondi congruence. 

On the other hand, the Sachs-Penrose reciprocity theorem,  states that:

\vspace{.2in}
\noindent{\bf Theorem}: {\it Assume that $X_1$ and $X_2$ are two matrices whose elements are the tetrad components of the two  complex deviation vectors associated with a null cone congruence with apex at a point $p_1$ and $p_2$ respectively, then    
$$
X_1 (\mbox { at } p_2)=-X_2 ( \mbox { at } p_1).
$$
 }

\vspace{.2in}
Hence using this theorem, we may assert that if the past null cone from $\scrip$ has a conjugate point at an affine distance $s=s*$ then the future null cone from $s*$ has a conjugate point at  $\scrip$. Then in the 
limit when $s_0 \to s*$, we find that $\rho_1(s_0,\scrip)$ goes to $+\infty$ and from(\ref{div}) $R(s*)=-\infty\Box$
\end{proof}

The first consequence of this lemma is that our coordinate system is well defined in the domain $R\in(-\infty,\infty)$, in other words from our coordinate system we can not detect the caustics that arise in the past null cones as we move into the space-time

The lemma is  also useful to answer the question we possed before, namely, if there exist asymptotically flat space times that can be covered with a global canonical coordinate system constructed from a single function $Z$. Using proposition 4.4.5 \cite{hawking} which states that any null congruence along a geodesic such that the affine length can be extended arbitrarily  has a pair of conjugate points and lemma \ref{lemma:R} we conclude that the coordinate system $(u, w,\bar w, R)$ derived from $Z$ cannot cover a space-time except for Minkowski space. We conclude that
\begin{proposition}
A single function $Z(x,\zeta, \zb)$  cannot generate the conformal structure of a radiative space-time except for the Minkowski space.
\end{proposition}

Is is easy to show that $\eth Z$ and $\ethbar Z$ remain finite at a conjugate point. This follows from the fact that both are constants along a null geodesic.

It is also of interest to analize the behaviour of the conformal factor $\Omega$, and $\Lambda=\eth^2Z$ near a caustic since they generate the underlying metric of the space-time. 

In a similar way than the  Lemma above, we consider the future null cone with apex at $s_0$. The Sachs Theorem tells us that at $\scrip$ 
\begin{equation}\label{Lambda}
\sigma_1(s_0,\scrip) = m^a m^b\nabla_a l_b =\sigma_B - \Lambda(s_0).
\end{equation}
where $\sigma_B$ is the shear   of a Bondi congruence. 

If the future null cone of $s_0$ has a conjugate point at $\scrip$, then the shear $\sigma_1(s_0,\scrip)$ is either $0$ or $\infty$. It then follows from the Sachs-Penrose reciprocity theorem and eq. (\ref{Lambda}) that $|\Lambda(s_0)|$ is either $0$ or $\infty$ if the past null cone from $\scrip$
has a conjugate point at $s_0$.

It remains to consider the conformal factor $$\Omega^2=g^{01}:=g^{ab}Z,_a\eth\ethbar Z,_b = \frac{dR}{ds}.$$

 Since $R(s)$ diverges as $s$ approaches a conjugate point while the affine length is a smooth non-vanishing function along the null geodesic it follows that $g^{01}$ also blows up at that point.

\section{Conclusions}

We have shown that our main variable can be regarded as the generating family $\widehat Z$ of a constrained Lagrange submanifold and that its level surfaces are constrained Legendre submanifolds that project down to past null cones from
$\scrip$. Furthermore, we have demonstrated that for a generic space-time this 
Lagrange submanifold starts diffeomorphic to the configuration space but it develops caustic sets, points on $T^*(M)$ where its projections are caustic points on $M$. Thus, except for Minkowski space, a single function $Z$ on configuration space does not give the conformal structure of the space-time.
At the caustic points $\eth \ethbar Z$ diverges. This means that the coordinate system constructed on the null cones is only locally defined but on the other hand one never sees the caustics since they are pushed out to $R = -\infty$.

Although the entire treatment so far has been kinematical we would like to think of our variable as coming from the solution of a set of field equations given on the space-time\cite{dynamics,i-k-r}.

It is clear from the previous results that the solution of those field equations must have multiple valuedness in order to generate the multiple branches needed to construct the generating family $\widehat Z$ of the Lagrange submanifold. These solutions  are defined in a 6-dimensional space, 4-spacetime coordinates and 2 parameters on the sphere, $(\zeta,\bar\zeta)$. We demand the solution to be globally defined respect to the parameters  $(\zeta,\bar\zeta)$, that is, it shall be a piecewise smooth function on the sphere (it could be multiple valued but always finite on the sphere). In this sence we say that the solutions shall be {\bf globally regular} on the space of parameters.

Alternatively, we could try to find field equations given on $T^*M$. In this case the solution would yield a global generating family $\widehat Z$ of a constrained Lagrange submanifold  that coincides with $Z$ in a neighbourhood of $\scrip$. This last approach will be further explored.

\subsection*{Acknowledgments}
We are  indebted
to  Ted. E. Newman and   Simonetta Frittelli  for many enlightening conversations
and suggestions. 
This research has been partially supported by AIT, CONICET and  CONICOR.

\end{document}